\documentclass[twocolumn]{aastex6}
\usepackage{amsmath}
\usepackage{txfonts}
\usepackage{natbib}
\usepackage{hyperref}
\usepackage{color}
\usepackage{enumerate}
\usepackage{listings}

\newcommand{\changed}[1]{#1}
\lstnewenvironment{sql}{\lstset{language=sql}}{}
\lstnewenvironment{Python}{\lstset{language=Python}}{}

\DeclareTextFontCommand{\l}{\ttfamily\hyphenchar\font=45\relax}

\begin{document}

\title{TANGOS: The Agile Numerical Galaxy Organization System}

\author{Andrew Pontzen$^1$}
\author{Michael Tremmel$^2$}
\affiliation{$^1$Department of Physics and Astronomy, University College
  London, Gower Street, London WC1E 6BT, UK \\
$^2$Yale Center for Astronomy \& Astrophysics, Physics
  Department, P.O. Box 208120, New Haven, CT 06520, USA}
\date{21 May 2018}

\begin{abstract}
  We present {\sc tangos}, a Python framework and web interface for
  database-driven analysis of numerical structure formation
  simulations.  To understand the role that such a tool can play,
  consider constructing a history for the absolute magnitude of each
  galaxy within a simulation. The magnitudes must first be calculated
  for all halos at all timesteps and then linked using a merger tree;
  folding the required information into a final analysis can entail
  significant effort. {\sc Tangos} is a generic solution to this
  information organization problem, aiming to free users from the
  details of data management. At the querying stage, our example of
  gathering properties over history is reduced to a few clicks or a
  simple, single-line Python command.  The framework is highly
  extensible; in particular, users are expected to define their own
  properties which {\sc tangos} will write into the database. A
  variety of parallelization options are available and the raw
  simulation data can be read using existing libraries such as {\sc
    pynbody} or {\sc yt}. Finally, {\sc tangos}-based databases and
  analysis pipelines can easily be shared with collaborators or the
  broader community to ensure reproducibility. User documentation is
  provided separately.
\end{abstract}

\section{Introduction}

Analyzing simulations of cosmological structure formation poses
a significant computational challenge. Large numbers of raw data points
must typically be reduced into scientifically relevant properties or
observables for each galaxy or halo. The resulting quantities must
then be interpreted, which often involves further processing --- for
example to see how a galaxy's properties vary over time. In this
paper, we introduce {\sc tangos}, a software package which aims to
make such processing painless.

The code has been developed over a decade, with roots in work
described by \citet{Pontzen08DLA}. That research had to collate
information about a large number of halos across a range of different
simulations (to piece together the way that galaxies are seen in
absorption against quasars). Two problems became apparent. First, our
cross-sections, column densities and other quantities were structured
as a series of files with increasingly obscure names and
relationships. Reading the results and, especially, combining
different outputs into a coherent analysis was slow and cumbersome.
Second, our large collection of scripts for performing calculations
all included similar ``boilerplate'' code. This boilerplate would open
a series of simulations, run through the halos within them, and write
out results. Even the simplest alteration (for example, adding a new
simulation) required copy-and-pasting changes to multiple source
files. The combination of these obstacles became a major impediment to
progress.

\begin{figure}
\includegraphics[width=\columnwidth]{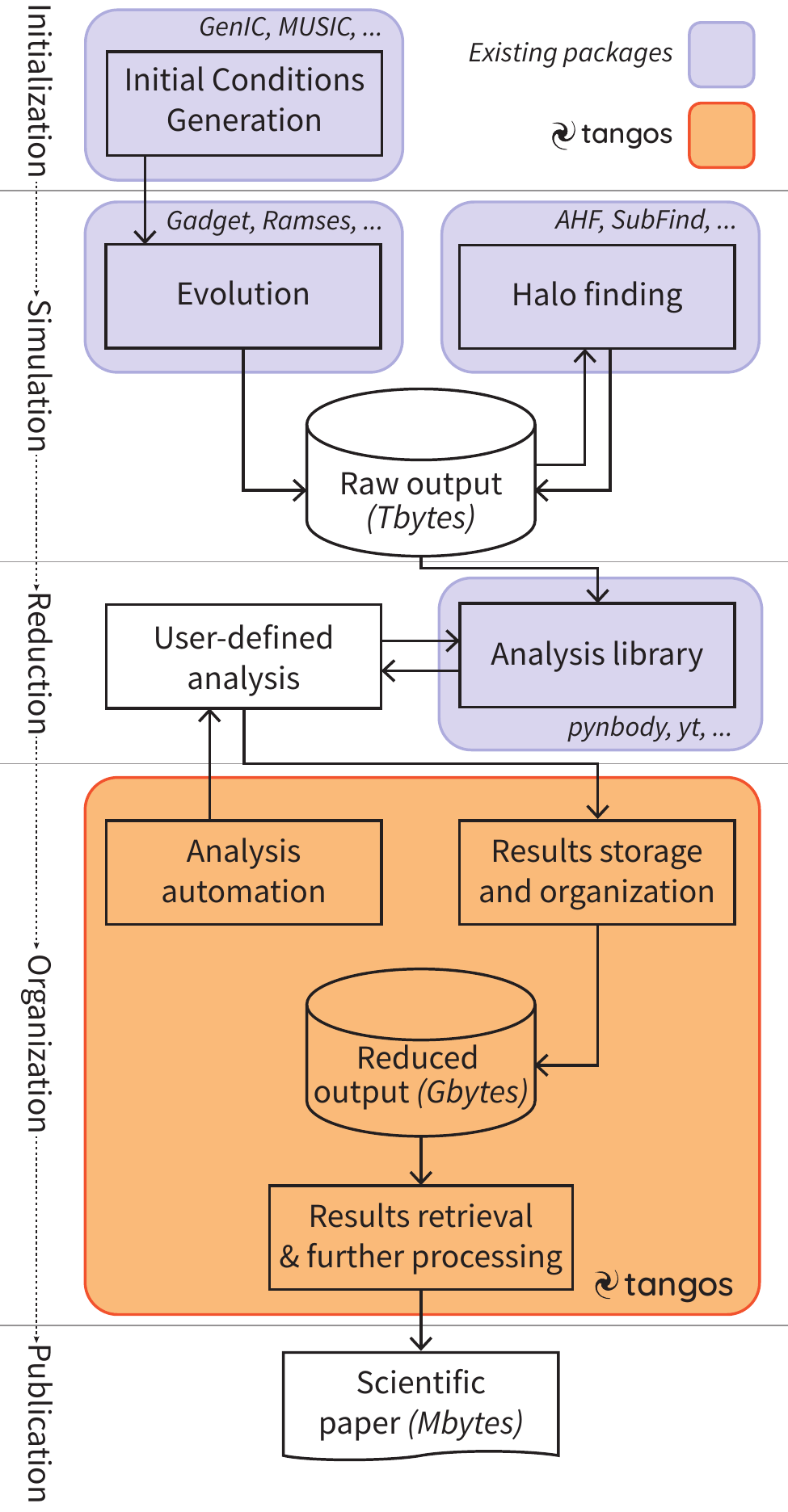}
\caption{An overview of how cosmological galaxy formation simulations
  are constructed and analysed. First, initial conditions must be
  generated with which to start the simulation. Then, the simulation
  is executed. After this, some form of data reduction is normally
  applied where the raw multi-terabyte outputs are distilled into
  scientifically meaningful quantities (for example, observable
  properties such as magnitudes and images or physical quantities such
  as masses and profiles). The fourth stage involves organising the
  data; {\sc tangos} is designed to take charge of this process,
  building an intermediate dataset that is typically gigabytes or
  even smaller. Finally, the results are retrieved in a form suitable
  for discussion in a scientific paper, further compressing the
  information to a point where it can be fully understood by a human
  reader; {\sc tangos} presents Python and web interfaces for this
  retrieval stage.}\label{fig:context-flowchart}
\end{figure}

Analyses of this sort suffer from simultaneously attempting to tackle
two conceptually separate problems: {\it reducing} the raw output to
scientifically-relevant quantities and {\it organising} the
results. Reduction and organization can be seen as two layers within a
simulation workflow (Figure \ref{fig:context-flowchart}).  By
separating the boilerplate organization layer, we started building a
generic code which would ultimately evolve into {\sc tangos}. The code
takes responsibility for storing and retrieving results as well as
iterating over simulations and halos to perform the reduction step on
all relevant data. Once this separation was made, we found we were
able to express science goals more clearly, leading to faster, higher
quality analyses.

We have been continually using and refining {\sc tangos} since that
time. In the last three years it has been heavily streamlined and
refactored to maximize the range of requirements it can accommodate
--- from traditional uniform volumes \citep[leading us to include
efficient parallelization, e.g.][]{Romulus17, diCintio17} to
``genetically modified'' zooms \citep[driving development of the
linking and tracking facilities;][]{Pontzen17GM}. To enable open
working with collaborators, we also added a web interface which can
formulate and process even complex queries.

In the meantime, codes such as {\sc yt} \citep{yt11} and {\sc pynbody}
\citep{pynbody13} have been maturing; these libraries present an
abstracted view of raw simulation data, aiding the reduction layer but
largely leaving organization to users. Both {\sc yt} and {\sc
  pynbody} contain some support for storing quantities such as
profiles alongside halo catalogues, but not for managing the results
of arbitrary user analysis. An add-on package for {\sc yt} known as
{\sc ytree} can generate and traverse merger trees but the user must
manually populate the data for each halo (unless using quantities
already calculated and stored by the halo finder, which are unlikely
to be sufficient for most use cases).  {\sc Halotools}
\citep{HaloTools17} is another existing code that addresses aspects of
an organization layer: it includes sub-packages to turn halo
catalogues into queryable merger trees. But its user tools are
focussed on constructing semi-analytic models from these trees rather
than populating them with properties calculated from the original
simulation data.

\changed{Conversely, the kind of questions that {\sc tangos} currently lends
  itself to answering center on hydrodynamic galaxy formation. How do star
  formation rates vary over time? What impact do mergers have on
  galaxy morphology? Where does a typical quasar line metal absorber
  lie relative to its nearest galaxy? Why does the distribution of dark
  matter get affected by some types of feedback but not others? By
  enabling many snapshots of multiple simulations to be linked together
  over time and efficiently queried, our ability to make progress in
  these areas has been enhanced.}

\begin{figure*}
\includegraphics[width=\textwidth]{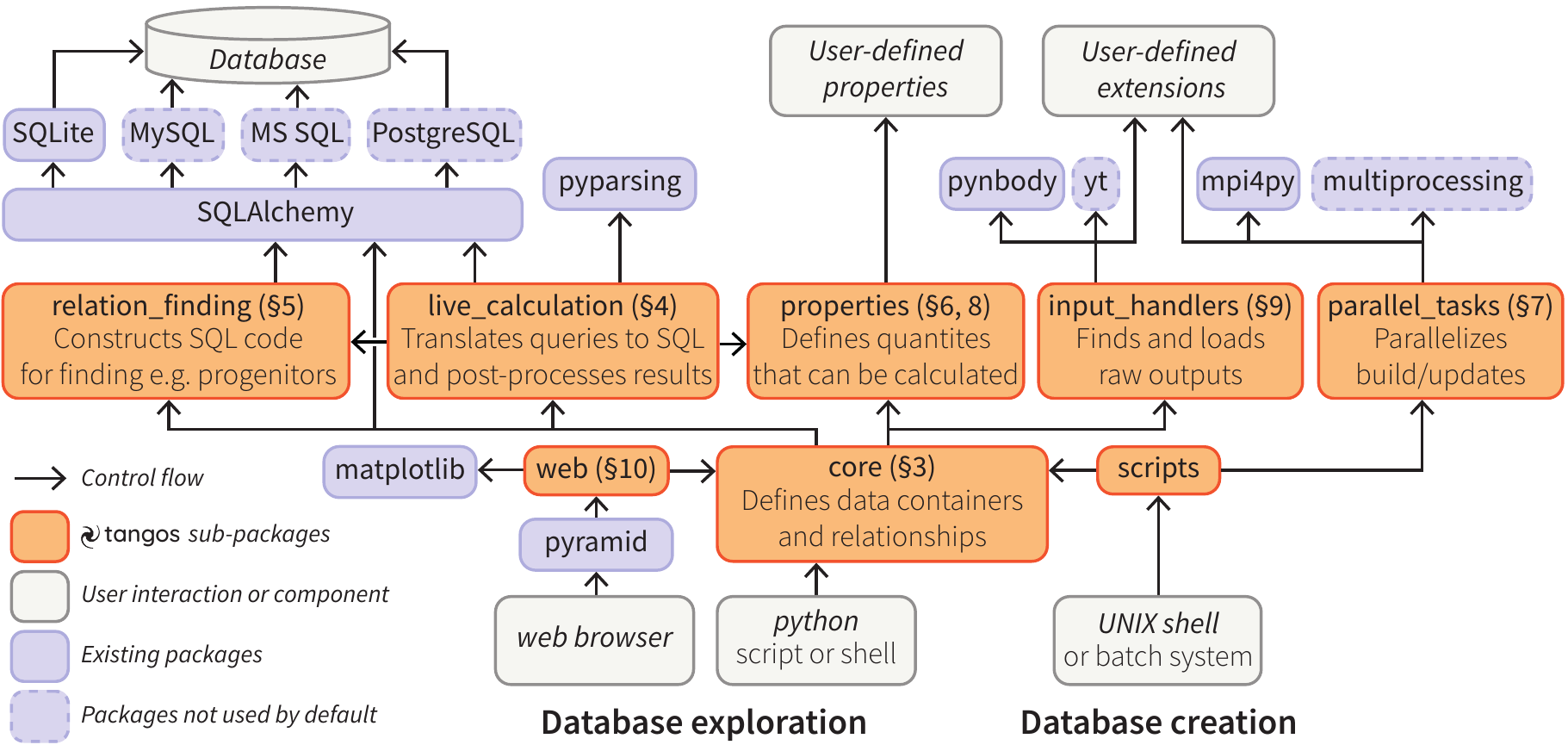}
\caption{An overview of the internal organization of {\sc tangos} and
  its interaction with external libraries and users. By partitioning
  responsibilities between various sub-packages, {\sc tangos} is made robust
  and flexible. Most modules are designed to be easily extensible. An
  overview is given in Section~\ref{sec:overview}.
}\label{fig:internal-organisation}
\end{figure*}

One way to conceive of a fully-fledged organization layer such as {\sc
  tangos} is from the perspective of data compression. Raw output from
the simulation layer can be extremely large (up to hundreds of
terabytes) because it contains snapshots of the full dynamical state
of the virtual universe at tens or hundreds of points in cosmological
time. The goal of analysis is to move from the multi-terabyte raw
output to a final, human-digestable set of information, nearly
always in the form of one or more scientific publications. To put an
upper limit on the information content of a paper, we can consider the
literal file size of the associated PDF (perhaps a few megabytes). The
compression ratio in moving from the raw output to the scientific
results is therefore $10^6$ or more. Inserting the organization layer
(giving two compression steps with ratios of order $10^3$) offers far
greater clarity and flexibility than attempting to jump six orders of
magnitude at once.

Some major simulation collaborations have made such intermediate
data public. Structured Query Language (SQL) databases
provided by the Millennium \citep{LemsonMillenniumSimDb}, MultiDark
\citep{Multidark}, Eagle \citep{McalpineEagleDb} and Theoretical
Astronomical Observatory \citep{BernykTAO} groups provide good
examples, as well as unstructured data releases with querying tools
like those provided by the Illustris collaboration
\citep{Nelson15Illustris}. However these tools are specific to
particular runs and pre-determined properties; they do not offer a
mechanism for adding new information or generating databases from
fresh simulations.

{\sc Tangos} instead aims to minimize the human effort required to
generate and collate complex results from new simulations of any type
and scale. By providing a framework that is modularized, {\sc tangos}
is extensible in multiple directions. Adding new galaxy properties,
querying techniques, parallelization methods, data storage approaches,
file formats and analysis libraries are all possible (and in some
cases, trivial).  The code is freely available \changed{from
  \url{github.com/pynbody/tangos} under an open source (BSD 3-clause)
  license} and compatible with Python 2.7 and 3.5 or
later. \changed{The version of the code used to prepare this paper is
  1.0.6 which is permanently available from Zenodo as
  \cite{tangos104}, although we would always advise using the latest
  available version for new projects.}

In this work we will discuss the overall structure of {\sc tangos} and
its modular components. Section~\ref{sec:overview} offers an overview
of the full system, while sub-components are explored in Sections
\ref{sec:schemaless-storage} to~\ref{sec:webserver}. We conclude in
Section~\ref{sec:conclusions}.

\section{Overview}\label{sec:overview}

{\sc Tangos} is implemented as a pure Python package and organized
into multiple sub-packages. The relationship between the various
sub-packages, user code and external dependencies is presented in
Figure~\ref{fig:internal-organisation}.  Practical documentation for using
 {\sc tangos}  can be found at \url{http://tiny.cc/tangos}.

At its core our solution consists of a storage engine implemented atop
{\sc SQLAlchemy}\footnote{\url{http://sqlalchemy.org}} which is a SQL
toolkit and object relational mapper. {\sc SQLAlchemy} does not
constitute a database in its own right, but rather presents a unified
high-level interface to multiple possible implementations such as
{\sc SQLite}, {\sc MySQL}, {\sc MS-SQL} and more. Our approach is
described in Section \ref{sec:schemaless-storage} and implemented in
the submodule \l{core}.

Querying the database can in principle be accomplished with raw SQL,
but it is easier to use our exposed {\sc SQLAlchemy} objects or
higher-level functions. To aid constructing queries, especially those
that traverse merger trees, we have implemented a domain-specific
mini-language which is parsed using {\sc
  pyparsing}\footnote{\url{http://pyparsing.wikispaces.com}} and
mapped into a combination of SQL and Python operations. This is
implemented in the \l{live\_calculation} module
(Section~\ref{sec:queries-without-sql}). The construction of SQL
queries for merger trees and other interrelationships between
objects is delegated to the \l{relation\_finding} module which is
described in Section \ref{sec:relation-finding}.

One of the strengths of the system is that users can easily define new
galaxy properties to be calculated at each stored timestep. The
framework allowing this extensibility is contained within the
\l{properties} sub-package and described in Section
\ref{sec:user-defin-analys}. When calculating properties, {\sc
  tangos}' \l{parallel\_tasks} sub-package enables parallelization via
several strategies which are described in Section
\ref{sec:parall-strat}.  We also allow rapidly time-varying quantities
such as star formation rates to be stored and processed at a
resolution finer than the snapshot steps, via a mechanism described in
Section \ref{sec:time-chunking}.

We factor out all tasks related to file
loading and memory management to the \l{input\_handlers} module which
is covered by Section \ref{sec:agnost-file-handl}. By extending this
module it is possible to work with any analysis toolkit for the
reduction layer, although the default is {\sc pynbody}.

The final component of {\sc tangos} is its web server which enables
databases to be explored from within a browser. The server, described
in Section \ref{sec:webserver}, does not implement any additional
functionality; it simply provides an alternative interface that is
convenient for rapid data exploration.

\section{The core}\label{sec:schemaless-storage}
\begin{figure*}
\begin{center}
\includegraphics[width=0.7\textwidth]{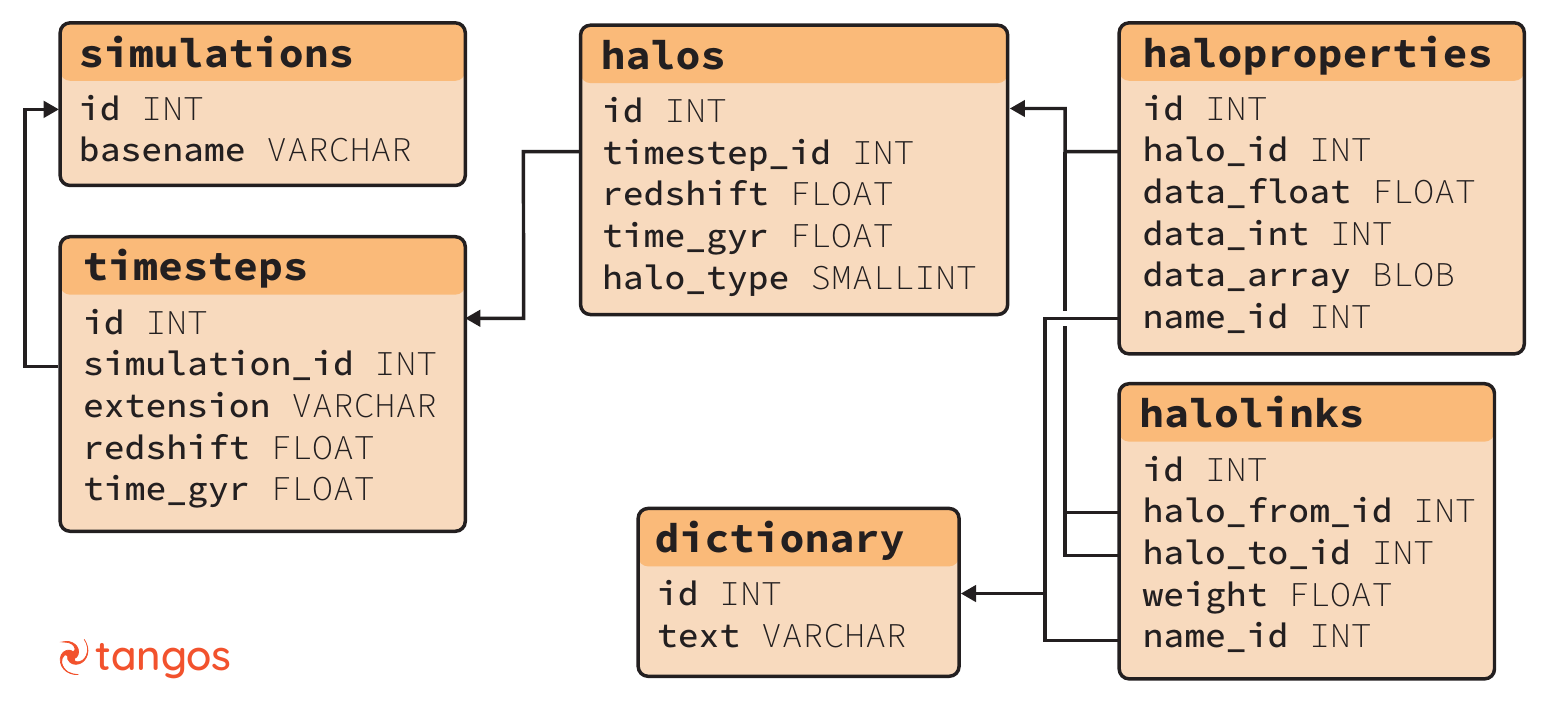}
\end{center}
\caption{The layout of the relational databases that {\sc tangos}
  builds. Simulations can be associated with any number of timesteps
  which in turn can be associated with any number of objects (halos,
  black holes, or tracker regions). The objects have properties
  and links associated with them, forming the basic elements of a
  scientific analysis. While users do not have to be aware of this
  structure, we show it here because its layout is
  substantially different from that of existing databases
  \citep[e.g.][]{LemsonMillenniumSimDb, Multidark}, reflecting the
  schemaless approach described in the text.}\label{fig:db-layout}
\end{figure*}

The \l{core} sub-package defines the layout of {\sc tangos} databases;
as we describe below this structure is not directly exposed to a
typical user. Our databases are relational, consisting of a number of
tables with pointers from one to another (Figure \ref{fig:db-layout});
our use of {\sc SQLAlchemy} allows the user to choose from a wide
variety of industry-standard underlying engines. By default we use
{\sc SQLite}\footnote{\url{https://www.sqlite.org}} which is a
serverless system: the database consists of a single file, so that it
can be downloaded from a cluster for offline analysis on a laptop or
other device\footnote{The UNIX tool {\sc rsync} is particularly
  suitable because {\sc SQLite} writes new entries to the end of
  existing files -- {\sc rsync} is efficient at then updating any
  local copies at minimal bandwidth cost. This enables a workflow
  where updated or new galaxy properties are remotely computed and a
  local database copy is kept up-to-date. For larger simulations and
  collaborations, however, it may be more practical to use {\sc
    tangos} with a client-server database system such as {\sc
    MySQL}.}. A number of indexes are created alongside the tables,
greatly enhancing the speed at which relevant queries may be executed at
the cost of slightly increased storage space for the resulting
database.

Each known simulation corresponds to a single entry in a
\l{simulations} table, linking to multiple entries in a \l{timesteps}
table, each of which in turn link to multiple entries in a \l{halos}
table. It is worth noting that the \l{halos} table can, in fact, store
multiple classes of object: halos, groups, tracked Lagrangian regions,
or individual black holes. Each of these objects behaves similarly
until the raw data is needed, at which point {\sc tangos}
automatically loads the appropriate portion.  Because of the need to
maintain backward compatibility during development, the table is still
known as \l{halos} but we refer to a generic entry in the table as an
{\it object}.

We associate any number of {\it properties} with each object. These
might be quantities such as magnitudes and masses or arrays such as
histograms and images. The properties are stored using a
``schemaless'' system, meaning that we do not create additional
columns for each property but rather link to entries in a
\l{haloproperties} table (with the name again reflecting a historical
choice). Schemaless storage systems are popular in industrial
applications (see
e.g. \l{MongoDB}\footnote{\url{http://www.mongodb.com}} and
\l{schemaless}\footnote{\url{https://eng.uber.com/schemaless-part-one/}}).
For {\sc tangos}, the primary advantage is one of simplicity in
managing the database: there is never any need to create or drop
columns.

Our approach effectively attaches key-value pairs to each object; the
keys are stored as a link to a separate \l{dictionary} table.  Because
querying is carried out through {\sc tangos} itself (Section
\ref{sec:queries-without-sql}), the user need not be aware of any of
these implementation details. \changed{Properties do not carry
  explicit units although this functionality is likely to be added in future.}

In addition to storing properties, objects can also be linked to one
another. Entries in a \l{halolinks} table describe this relationship
between two objects; for example, one halo might be a progenitor,
descendant, or subhalo of the other. We will discuss links and how
they are used to generate informative science queries in Section
\ref{sec:relation-finding}, but first we consider the more elementary
retrieval of properties from one or more objects.

\section{Queries and calculations}\label{sec:queries-without-sql}

The recommended approach to querying a {\sc tangos} database is
through the provided Python or web interfaces. Basic queries can be
executed through a dictionary-like syntax. For example:
\begin{Python}
sim = tangos.get_simulation('my_simulation')
timestep = sim[42]
halo = timestep[5]
print(halo['V_mag'])
\end{Python}
will access the stored $V$ magnitude of halo $5$ in the $42$nd output of
\l{my\_simulation}.  Each line of Python code is translated by {\sc
  tangos} into a {\sc SQLAlchemy} query, which in turn emits the correct
dialect of SQL and returns the result.

This approach is acceptable for interactive exploration of small
amounts of data. However when larger quantities of data are to be
retrieved, issuing a series of multiple small queries is an
inefficient approach since there is substantial latency associated
with each round trip to the database. It is more time-effective to
retrieve all required data in a single query. {\sc tangos} offers
multiple routes to optimizations of this sort.

For example, it is possible to retrieve properties from a series of
objects, such as all those in a timestep. It is equally possible to
retrieve multiple properties from each object.  Combining both, the
query
\begin{Python}
B, V = timestep.calculate_all("B_mag","V_mag")
\end{Python}
returns two {\sc numpy} arrays with respectively the B- and V-band
magnitudes of every halo in the timestep. To achieve this, {\sc
  tangos}  generates and executes optimized SQL consisting of a
double join from \l{timesteps} to \l{halos} and on to
\l{haloproperties}.

We additionally implemented a mini-language to enable
calculation within queries. The code
\begin{Python}
color = timestep.calculate_all("B_mag - V_mag")
\end{Python}
results in the SQL join described above, but a post-processing step in
Python takes the column difference before returning the resulting
array to the user.  While this simple example could be fully executed
in SQL, the hybrid SQL--Python approach allows for more complex
expressions. Users can even define functions which may be used within
queries (see Section \ref{sec:user-defin-analys}); a built-in example
is the \l{at} function. The request
\begin{Python}
rho = timestep.\
 calculate_all("at(Rvir/2, dm_density_profile)") 
\end{Python}
returns the value of the dark matter density profile evaluated at half
the virial radius for each halo. Within \l{calculate\_all}, the
evaluation takes place in stages. First the user's mini-language
string is parsed and turned into an abstract syntax tree; by
inspecting this tree, we can identify \l{Rvir} and
\l{dm\_density\_profile} as the underlying properties to be
retrieved\footnote{User-defined functions can also demand access to
  properties that are not explicitly referenced in the query tree;
  these properties are included in the join. See Section
  \ref{sec:user-defin-analys}.}. Next, the SQL is generated and
emitted. Finally, the appropriate Python functions are called: a {\sc
  numpy}-implemented divide operation followed by the interpolation
function \l{at} which generates the final result.

The query system allows access to linked objects' properties such as
those from a halo's major progenitor $n$ timesteps earlier. One might
be interested in how much each galaxy's magnitude has changed over
the last $n=5$ steps, which corresponds to the query
\begin{Python}
dV = timestep.\
      calculate_all("V_mag - earlier(5).V_mag")
\end{Python}
Such calculations involve joins onto the requested properties from
a heterogeneous set of halos which must first be
identified using a merger tree. This is implemented within the \l{relation\_finding}
sub-package, as we now describe.

\section{Merger trees and other relationships}\label{sec:relation-finding}

Understanding how galaxies change over time is at the heart of many
science analyses. This requires {\sc tangos} to store and query merger
trees, which express the hierarchical merging of structures
\cite[e.g.][]{KauffmannWhite93}.  We may additionally be interested in
non-temporal relationships such as whether a particular halo is a
subhalo of another, or to which halo a black hole is associated. As a
final example, it can be useful to provide a map from halos in one
simulation to those in another (given closely related simulations
based on the same initial conditions). {\sc Tangos} addresses all
these needs by storing {\it links} between objects. Links are unidirectional
and come with an associated {\it weight} which determines the strength
of the relationship in a way to be defined shortly. The links are
stored in an underlying table named \l{halolinks}. (As a reminder, the
historically-chosen name belies that links do not have to connect
halos: they can point from any object to any other.)

Each connection in a merger tree corresponds to two links pointing
respectively forwards and backwards in time. This allows us to define
independent weights for each direction by the number of particles
in common as a fraction of the number in the link source. For a 
halo merging into a larger structure the forward weight
will be close to $100\,\%$ whereas the reverse weight will give the
merger ratio. The \l{relation\_finding} sub-package is able to use
this information to respond to queries as follows.

The sub-package assigns every halo a \l{previous} and \l{next}
property such that the Python code
\begin{Python}
halo_prog = halo.previous
\end{Python}
computes the major progenitor of \l{halo} in the previous
timestep. When accessed by a user, \l{previous} generates and
processes a suitable joined SQL query between the \l{halos} and
\l{halolink} tables. If more than one linked halo is available, the
link with the highest weight is selected (since other links
point to smaller merging structures). The \l{next} property
operates in a similar way.

\begin{figure}
\begin{center}
\includegraphics[width=0.95\columnwidth]{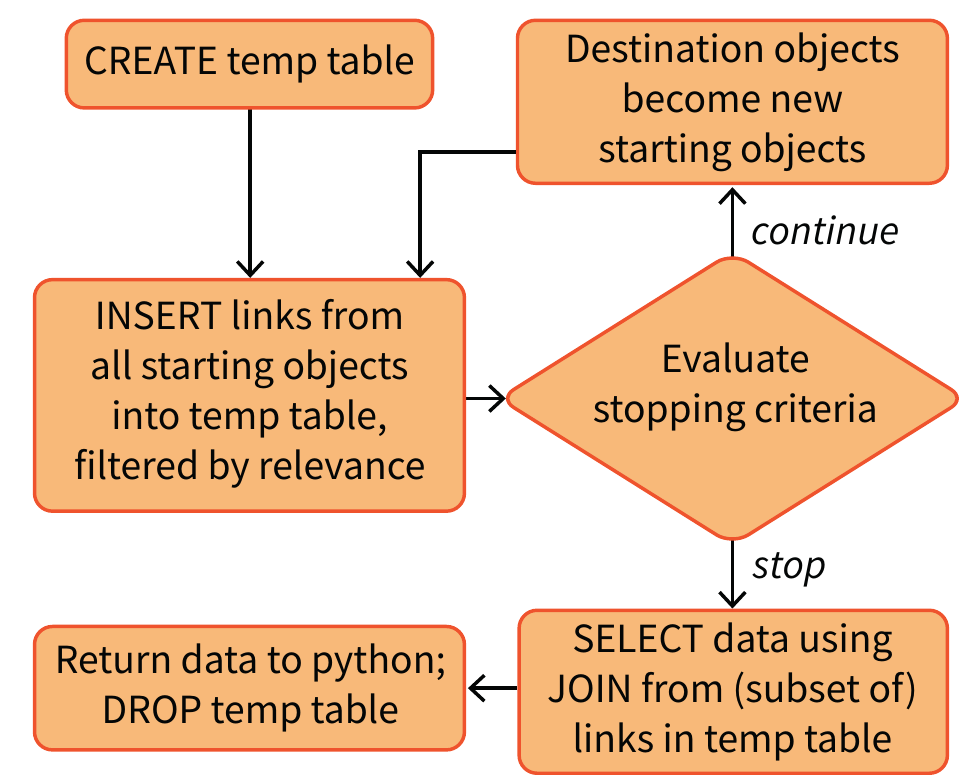}
\end{center}
\caption{The algorithm used to find relations such as major
  progenitors in a specified timestep. The database holds links only
  between adjacent timesteps, but by recursively populating a SQL
  temporary table, {\sc tangos} can follow these links across multiple
  timesteps in an efficient way. User-requested data from the
  progenitors is acquired by a join onto the temporary table and
  delivered to Python at the end of the recursive process.
}\label{fig:tree-search}
\end{figure}

While this is the simplest example of using links, there are
considerably more powerful options available when \mbox{multiple} halos or
timesteps are involved.  We have implemented a series of {\it
  strategies} that efficiently find halos in several such scenarios.
One typical use case is to collect properties along an entire
major progenitor branch. To collate the color discussed in Section
\ref{sec:queries-without-sql}, we use the request
\begin{Python}
cols = halo.calculate_for_progenitors('B_mag-V_mag')
\end{Python}
which will be executed in multiple stages within the database (Figure
\ref{fig:tree-search}), followed by final processing in Python. The
approach is to create a temporary table (which in most SQL
implementations is possible without write access to the database)
mirroring the structure of the \l{halolinks} table.  It is first
populated by inserting the links leading directly away from the
initial objects. However only links satisfying a relevance criterion
are included; for example, in the major progenitor search, they must
point to an earlier timestep and have a weight higher than any other
link to that timestep. Our implementation allows the filter to be
redefined for different use cases (see below).  The system next
evaluates a stopping criterion against the links in the table; in the
case of our progenitor search, it simply checks whether any new
objects were uncovered in the most recent cycle. If so, the procedure
starts again (now searching for links from the most recently
discovered objects). If not, the recursion is complete.

By design, no data is transferred from SQL to Python during the
recursion since this would be needlessly slow. Instead, at the end of
the loop, the temporary table is handed to the \l{live\_calculation}
module. That system prepares a query against the temporary table
joined to \l{haloproperties} as described in Section
\ref{sec:queries-without-sql}. The resulting columns are then
retrieved and processed; the temporary table is dropped; and the
results are returned.

The basic algorithm has been customized for a wide range of scenarios
by subclassing the base \l{MultiHopStrategy}. In addition to the major
progenitor search described so far, we have implemented subclasses
that search for:
\begin{enumerate}[(i)]
\item all progenitors (rather than major progenitors) -- this 
  required us to remove the highest-weight restriction in the
  definition of relevant links;
\item major descendants -- accomplished by reversing the timestep comparison;
\item the major progenitor or descendant in a particular timestep
  (rather than all timesteps) -- achieved by stopping once objects
  from that timestep are discovered, and selecting only those objects;
\item corresponding halos -- defining relevant objects as those in
  another simulation at the same physical time, and stopping once such
  an object is discovered;
\item the most recent merger -- similar to the all-progenitor search,
  but with a stopping criteria that halts when more than one
  progenitor is found in a single timestep.
\end{enumerate}

Users typically access these strategies through the
live calculation mini-language.  Consider the query given at the end
of Section \ref{sec:queries-without-sql}:
\begin{Python}
dV = timestep.\
      calculate_all("V_mag - earlier(5).V_mag")
\end{Python}
which requests, for each halo in \l{timestep}, the difference between
the present $V$ magnitude and the major progenitor's $V$ magnitude 5
timesteps earlier. For these purposes, strategy (iii) in the list
above is applied. Use of other strategies is described in the documentation.

\begin{figure*}
\begin{center}
\includegraphics[width=0.9\textwidth]{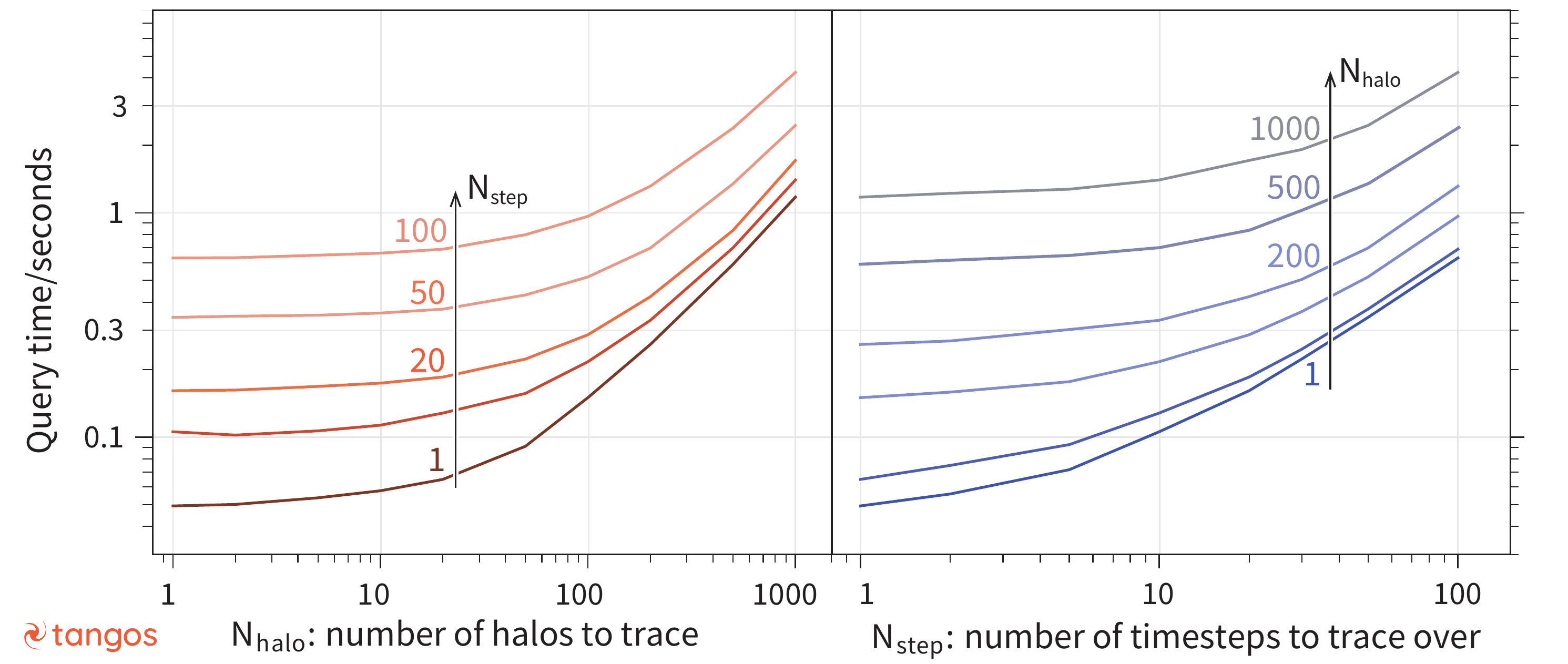}
\end{center}
\caption{Performance for querying properties of the major progenitor
  of $N_{\mathrm{halo}}$ halos, $N_{\mathrm{step}}$ steps prior to a
  selected snapshot. Both panels show the same results, cutting across
  the 2D $(N_{\mathrm{halo}}, N_{\mathrm{step}})$ plane in
  complementary ways. The time shown is the end-to-end {\sc tangos} query
  runtime on a {\sc SQLite} database for a uniform resolution
  simulation on a 2013 Macbook Pro (Intel Core i7-4850HQ 2.3 GHz with
  16GB RAM). The runtimes are insensitive to the number of properties
  queried since the majority of time is spent in the
  \l{relation\_finding} sub-package.}\label{fig:performance}
\end{figure*}

Taken together, the \l{relation\_finding} and \l{live\_calculation}
modules allow {\sc tangos} to execute queries that would be
exceptionally hard to express within native SQL and prohibitively slow
to implement in pure Python.  Performance of the resulting system is
explored in Figure \ref{fig:performance} for a {\sc SQLite}-backed
database of a uniform volume simulation. Querying was undertaken
on a 2013 Macbook Pro (Intel Core i7-4850HQ 2.3 GHz with 16GB RAM)
running Python 3.6.1 and {\sc SQLite} 3.13.0. We used a simulation with
$100$ steps spaced equally in time between $z=1$ and $z=0$. This
is unusually fine time spacing, but allows us to explore the
performance implications of tracking large numbers of objects over
many steps.

The panels of Figure \ref{fig:performance} plot the same information
projected along different axes. The two variables $N_{\mathrm{halo}}$
and $N_{\mathrm{step}}$ represent the number of halos to trace and of
timesteps to jump. (We found that the number of properties ultimately
retrieved from each halo is irrelevant since the time for computing
progenitors is always the limiting factor.)  As either
$N_{\mathrm{step}}$ or $N_{\mathrm{halo}}$ becomes sufficiently large
(exceeding $20$ \changed{or $100$ respectively}) the scaling is approximately
linear. The shallow scaling at smaller $N$ reflects the efficiency
savings to be found in minimizing the number of SQL queries
generated. Since the $N_{\mathrm{step}}=1$, $N_{\mathrm{halo}}=1$ case
takes approximately \changed{$0.05$} seconds to complete, our most
ambitious $N_{\mathrm{step}}=100$, $N_{\mathrm{halo}}=1000$ case would
take \changed{well over an hour} without optimization (compared to
\changed{4.2 seconds} with the optimizations, \changed{over a
  thousand-fold improvement}).

Bearing in mind that {\sc tangos} sees the user's time and patience as
a limiting resource, this is a major enhancement. It enables more
typical queries -- say with $1000$ halos and $20$ timesteps -- to
complete in under two seconds. \changed{In this case, $320\,$ms is
  spent on building the temporary table which follows 1000 halos back
  through the merger tree; $1200\,$ms is spent filtering the completed
  temporary table and gathering the requested data; and $110\,$ms is
  spent on post-processing. The first two steps ($1520\,$ms) are
  carried out within the database engine, while the last ($110\,$ms)
  is executed by Python. }

\changed{For some cosmological applications it may be useful to
  address a significantly larger number of halos; we can
  extrapolate the performance to query times of ten minutes for
  tracing $10^6$ halos over $20$ timesteps, although we did not
  explicitly test this. Improving run-times in this case would most
  likely be possible by replacing {\sc SQLite} with a server-based
  database engine. }




    





\section{User-defined properties}\label{sec:user-defin-analys}

Adapting {\sc tangos} to a given science case requires the user to
develop a set of {\it properties} which will be calculated for each
object within the database.  A new property is defined by implementing
a subclass of \l{PropertyCalculation}. At a minimum, the property
author must override \l{names} (which specifies one or more names of
the properties to be stored) and \l{calculate} (which computes the
values of those properties). \changed{The framework is responsible for
passing raw data to \l{calculate} and for writing the
returned value into the database.} Therefore property classes express
only scientific intent and do not take any responsibility for I/O.

Properties may either be dependent on existing entries in the
database, or on the raw simulation snapshot data, or on both. For
example, a halo virial velocity can be derived from an existing
measurement of the virial mass but the calculation of a dark matter
density profile would likely need access to the snapshot data. By
default, any raw data provided to a calculation includes only the
particles or cells within the object under consideration. However it
is possible to provide a \l{region\_specification} method to request
access to a more extended volume; one use we have made of this
facility is to measure inflows and outflows across the virial radius.

{\sc Tangos} provides two routes to performing calculations: through a
command-line script (\l{tangos write}) which writes results into the
database, and through the \l{live\_calculation} system which
calculates on-the-fly during a query and does not store any output
(Section \ref{sec:queries-without-sql}). Properties requiring raw
simulation data are only available when using the command-line tool.
As we will discuss in the next section, the \l{tangos write} script
can be parallelized at the halo or timestep level. Because the
property class contains no explicit I/O, the implementer does not
normally have to plan for different parallelization scenarios and can
instead focus on science goals.

\section{Parallelization strategies}\label{sec:parall-strat}

{\sc Tangos} offers a parallelization scheme, implemented by
\l{parallel\_tasks}, for use when building or updating databases. It
is primarily targeted towards systems with a Message Passing
Interface (MPI) library available, although it can also make use of
Python's \l{multiprocessing} module in place of MPI if necessary.

Given that analysis typically consumes a small fraction of the total
computing resources of a simulation, our focus in implementing
parallelization is on user convenience rather than machine
efficiency. Nonetheless reduction of simulation data to a set of
properties is a near-perfectly parallel process since each halo (or,
potentially, timestep) can be considered independently of all
others. This independence allows {\sc tangos} to offer multiple
parallelization options, each with differing benefits. Users
select from these options at \l{tangos write} runtime by specifying a
{\it load mode}. 

In practice any given load mode is reliant on the input handler
(Section \ref{sec:agnost-file-handl}) which is responsible for
providing data as the parallel calculation proceeds.  Here, we
describe the parallel capabilities of the default {\sc pynbody} input
handler which implements four modes.  The user starts \l{tangos write}
through \l{mpirun} which launches multiple processes; all load modes
use the first of these as a server with differing responsibilities as
follows.

In the \l{default} load mode (which is applied if the user does not
specify an alternate), the server assigns entire timesteps to all
other processes. The cores then operate independently on each
snapshot, running through the requested calculations for relevant
halos and other objects. The chief advantage of this approach is
simplicity and lack of communication overheads. However, it can lead
to problematically large demands on memory: the number of cores used
per node will need to be manually reduced using an appropriate
invocation of \l{mpirun} if the size of a snapshot $M_{\mathrm{snap}}$
is too large (i.e. if $NM_{\mathrm{snap}}>M_{\mathrm{node}}$ where
$M_{\mathrm{node}}$ is the total RAM available per node and $N$ is the
number of MPI processes per node). Reducing $N$ may anyway be
desirable if the property calculations are themselves parallelized
using threads (see below).

When the \l{partial} load mode is specified, the server operates
similarly but each individual core activates {\sc pynbody}'s {\it
  partial loading}. This avoids the full simulation snapshot being
retrieved from disk, instead loading the data for a single halo at a
time. Partial loading is a simple solution to reducing memory usage,
but can lead to excessive disk access (especially for network file
systems) as multiple processes may simultaneously access data spread
in near-random patterns across large files. It is also not advisable
to load particle data from custom regions (Section
\ref{sec:user-defin-analys}) in this way, since currently such
requests to {\sc pynbody} can only be satisfied by scanning the
properties of all particles in the snapshot.

The \l{server} load mode addresses these shortcomings by taking an
entirely different approach. The server process takes responsibility
for loading an entire snapshot, and then instructs the worker
processes to perform calculations for individual halos or other
objects. These cores respond with a request (or possibly multiple
requests) for the raw data required\footnote{This occurs transparently
  from the point of view of user analysis code. The actual requests
  for data take place through {\sc pynbody}'s lazy-loading
  mechanism.}. The chief advantage of this approach compared with
partial loading is that disk access is consolidated into a larger,
sequential read. The peak memory usage of server mode is only slightly
greater than $M_{\mathrm{snap}}$, and this peak only occurs on the
first node. In systems with heterogeneous hardware, it is possible to
assign the server process to a machine with expanded RAM (e.g. the
\l{bigmem} nodes on NASA's Pleiades) while using regular nodes for the
calculations. If the memory cost of loading an entire snapshot is
still prohibitive, a hybrid \l{server-partial} mode loads only minimal
information such as particle positions rather than the entire snapshot
on the server. The individual worker nodes then load the relevant
portions of all other arrays through the partial-loading approach.

One drawback of \l{server} mode is that it can generate
significant MPI traffic to remote nodes which continually require
new data.  A final alternative, \l{multi-server}, is currently being
planned where each node runs its own local server process. Provided
every node has $M_{\mathrm{machine}}>M_{\mathrm{snap}}$ this approach
should offer greater efficiency by minimizing network traffic.

In tandem with all the above, the user can implement their own
per-halo shared-memory parallelization (based on threads or {\sc
  OpenMP} from {\sc cython}\footnote{\url{http://cython.org}}) and reduce the number
of MPI processes spawned to free up physical processing cores. This
approach to parallelization can be highly efficient but will typically
require effort for the property implementer, unless the calculation
is chiefly carried out by library routines. Luckily many {\sc pynbody} built-in
routines (such as smoothing and image rendering) are already
parallelized with threads. The optimal balance of threads and MPI
processes for a particular calculation can be determined empirically
if required.

\section{Sub-step time resolution}\label{sec:time-chunking}

In Section \ref{sec:user-defin-analys}, we discussed how new
properties can be associated with objects at each timestep; their
variation over time can then be inspected using the algorithms from
Section~\ref{sec:relation-finding}. However, this ties the time
resolution of stored properties to the simulation's snapshot interval
whereas many analyses of galaxy formation benefit from studying
more rapid variations. The star formation or
black hole accretion rates can vary by orders of magnitude over even a
small fraction of a galaxy's dynamical time. It is infeasible to store
and process snapshots at sufficiently regular  intervals to
capture such short timescales.

\begin{figure}
\begin{center}
\includegraphics[width=0.95\columnwidth]{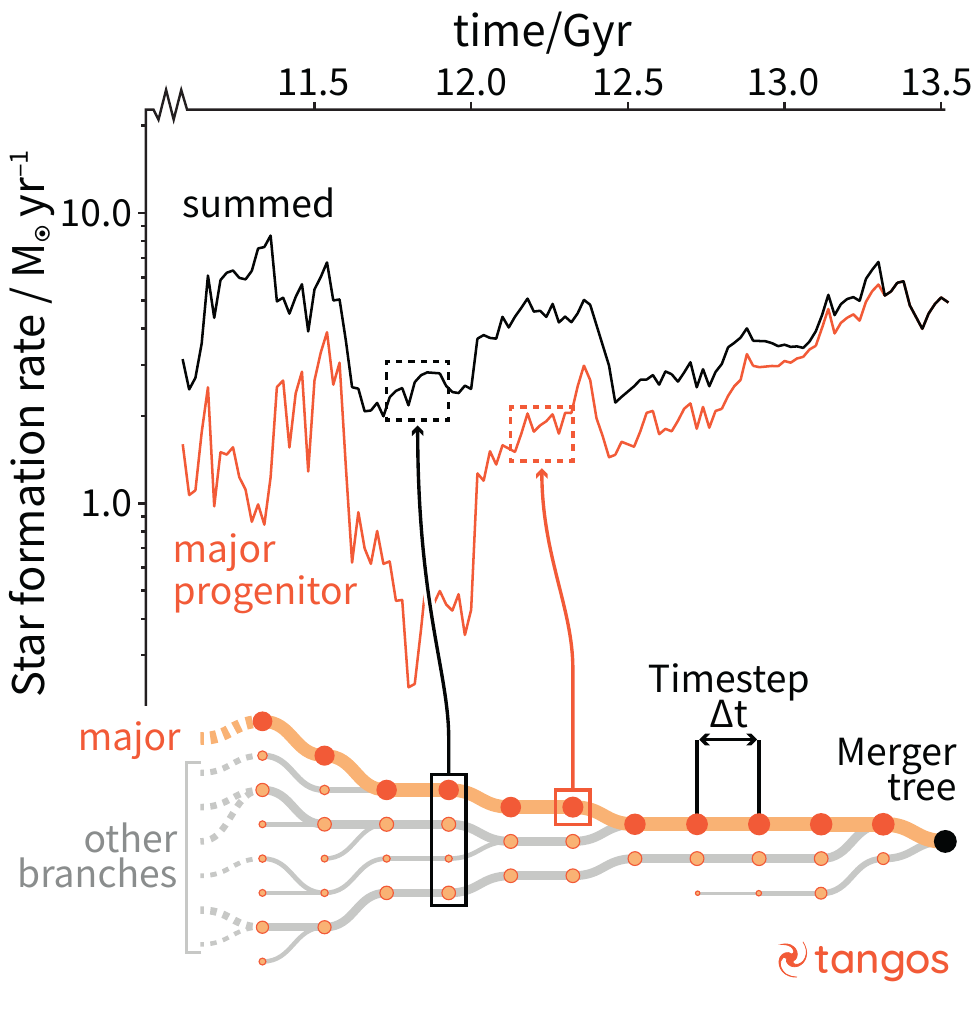}
\vspace*{-0.5cm}
\end{center}
\caption{The mechanism for storing and reconstructing time series with
  finer-than-timestep resolution, illustrated for a star formation
  rate history. Each halo stores a small chunk of history that covers
  the time back to the previous stored step. When the star formation rate
  is retrieved, {\sc tangos} automatically reassembles the individual
  chunks into a complete high-resolution history. The user can
  control whether to include only the major progenitor branch
  or, conversely, sum over all branches. }\label{fig:time-chunking}
\end{figure}

Simulation codes can work around this problem by storing auxiliary
data such as formation times for each star particle or a black hole
accretion log. {\sc Tangos} incorporates a general mechanism to
organize this information into {\it time chunks}, illustrated in
Figure \ref{fig:time-chunking}. In our scheme, the recent history of
star formation or accretion is stored with each halo, black hole, or
other object in each timestep. When the user attempts to access a
time-chunked property, {\it reassembly} is automatically initiated: by
default, {\sc tangos} finds all major progenitors of the object and
retrieves a chunk from each. It then constructs a history by gluing
the chunks together (orange line, Figure~\ref{fig:time-chunking}).
This approach allows significant flexibility in the manner of
reassembly. For example, the user can request chunks to be summed over
{\it all} progenitors (black line in Figure~\ref{fig:time-chunking})
rather than following the major progenitor branch.

From the property implementer's perspective, using time chunking is
straight-forward; instead of deriving from \l{PropertyCalculation} we
derive from \l{TimeChunkedPropertyCalculation} which implements the
reassembly discussed above. From a database user's perspective, the
mechanism is almost totally transparent because the reassembly process
is triggered by any request for the property. Switching modes to
obtain a summed history requires use of the mini-language
function \l{reassemble}, which is further described in the user
documentation.

\section{File handling}\label{sec:agnost-file-handl} 

We discussed in Section \ref{sec:user-defin-analys} how the \l{tangos
  write} tool loads data from a raw simulation and provides it to
users' property calculations. In addition, a \l{load} method is
associated with each object; this allows raw data to be loaded into a
standard Python session. Both approaches are implemented by the
sub-package \l{input\_handlers} which by default uses
the {\sc pynbody} library to load the underlying
snapshot. We have been careful to isolate {\sc pynbody}
references to within a single class which is used only if required
(specifically \l{PynbodyInputHandler}\footnote{Additionally,
  parallelization support is provided by the
  \l{parallel\_tasks.pynbody\_server} module which is
  loaded on demand.}) so that the system can be fully decoupled; for
example, an alternative \l{YtInputHandler} is provided to use the {\sc
  yt} library in {\sc pynbody}'s place. Users can straightforwardly
re-implement this functionality using different libraries if required.

In addition to loading the raw data on demand, input handlers take
responsibility for a diverse range of operations such as searching the
file system for available snapshots and enumerating the halos and other
objects within those snapshots. All these operations are therefore
user-customizable. Handlers are implemented by deriving a class
from \l{HandlerBase} and overriding a few methods to provide the
required functionality. In particular, the data returned from an
object's \l{load} method and passed to \l{PropertyCalculation}
instances is simply that returned from the underlying handler's
\l{load\_object} method. A custom handler class can be specified when
adding the simulation to a {\sc tangos} database (through the
\l{tangos add} script); it is then automatically used for all future
operations requiring data from that particular simulation. 

Adapting an existing input handler is also straight forward; one can
derive from an existing class (such as \l{PynbodyInputHandler}) and
override only those functions which need customization. The current
version of {\sc tangos} includes minor adaptations for working with
different file formats; for example, when working with {\sc SubFind}
catalogues, it is helpful to make distinctions between groups and
halos that do not necessarily exist with other halo finders. Further
information can be found in the user documentation.

\section{Web server}\label{sec:webserver}

\begin{figure*}
\begin{center}
\includegraphics[width=0.9\textwidth]{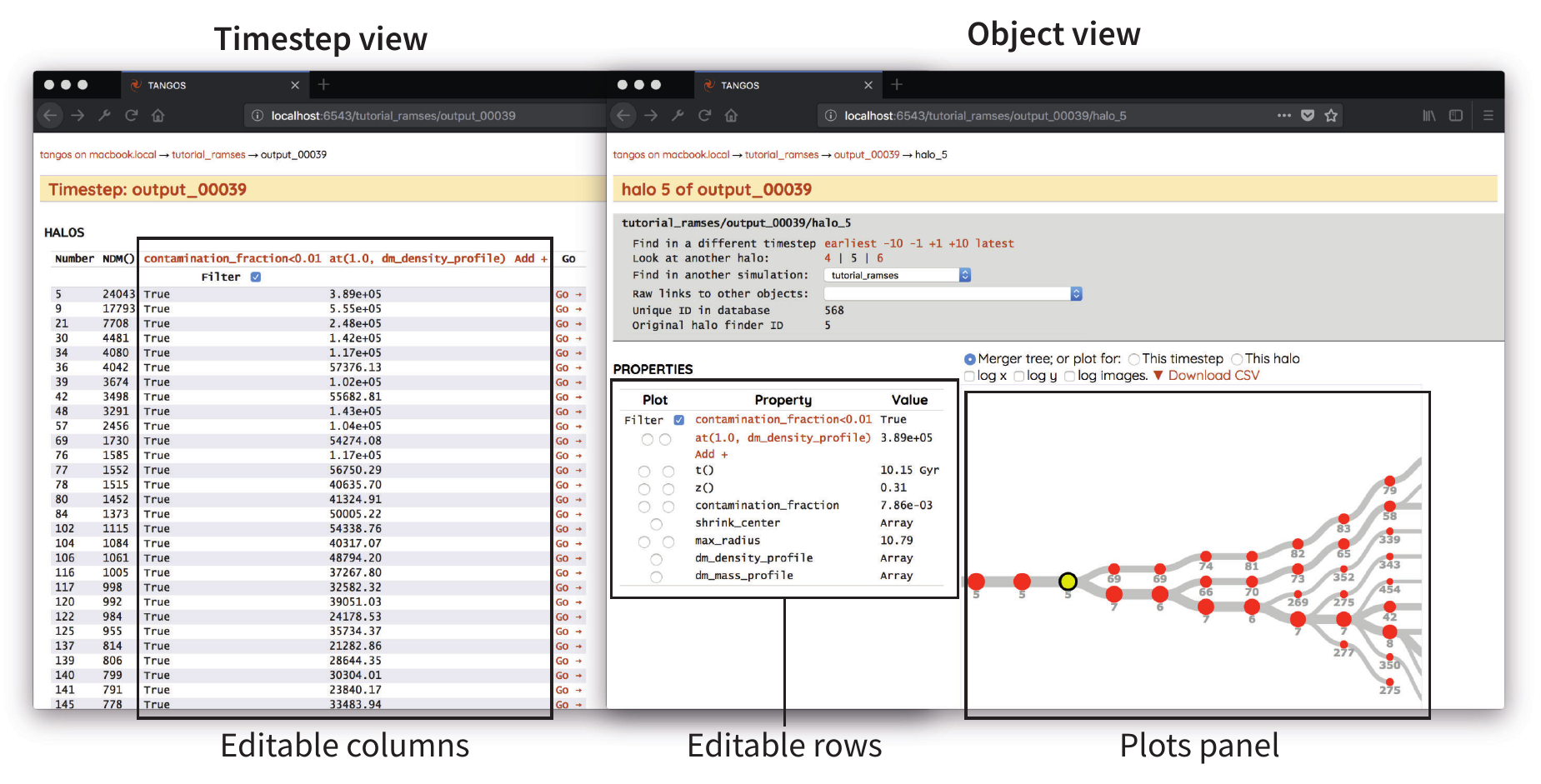}
\vspace{-0.3cm}
\end{center}
\caption{Two example pages from the web server. ({\it Left}) the
  timestep view, displaying one row per object and customizable
  columns which can be added, edited or removed and accept input in the
  live calculation mini-language for interactive queries. ({\it
    Right}) the halo view, which also contains a plots panel,
  here configured to show the halo merger tree. The rows are
  again editable, allowing for queries to be constructed and
  plotted from within a browser. }\label{fig:web}
\end{figure*}

To enable rapid exploration of the database by users, collaborators
and the broader community, {\sc tangos} includes a web server built
with the {\sc pyramid}\footnote{\url{http://trypyramid.com}}
framework. Typically a user will launch the server
application on their own machine and connect to
\l{localhost}. As a safe default, our {\sc pyramid} setup will not
accept connections from external machines (although it can be
tunnelled through SSH to a remote analysis node). If desired
{\sc pyramid} can be installed on a server and made world-accessible
through a single change to the configuration file.

The pages served by {\sc tangos} follow a natural hierarchy: the front page contains a list
of known simulations, with links to subsequent pages listing timesteps
and then objects.  In Figure \ref{fig:web} we show the latter two
stages using screenshots from a tutorial video\footnote{See
  \url{http://tiny.cc/tangos}.}. The left panel shows the timestep
view. In addition to basic information about each halo the user may
add any number of columns using the mini-language described in Section
\ref{sec:queries-without-sql}; as usual, this is parsed and executed
by the \l{live\_calculation} module. Here, for example, a query to
determine whether the halo is in the high-resolution portion of a zoom
simulation has been added, as well as a query which retrieves the dark
matter density at $1.0\,\mathrm{kpc}$. 

The right panel of Figure \ref{fig:web} shows the object view.  It
contains editable rows which correspond to the columns in the timestep
view. Additionally it can be used to display a variety of plots; in
this instance an interactive graphical representation of the merger
tree has been generated. This is handled by the
\l{relation\_finding} sub-package, which uses the all-progenitor
strategy (Section \ref{sec:relation-finding}) to trace the tree
(pruned by user-defined criteria such as minimum mass ratios).

Enabling interactivity requires part of the \l{web} module to be
written in {JavaScript} and executed within the browser. When the
user interacts with a page, the {JavaScript} code places
asynchronous requests to Python over HTTP. The server
calls the relevant functionality within {\sc tangos} and encodes the
results into JSON ({JavaScript} Object Notation); once the results
arrive back in the browser, {JavaScript} places them into the
appropriate elements within the page. Plots are generated using the
{\sc matplotlib}\footnote{\url{https://matplotlib.org}} library and
returned as a PNG file, with the exception of merger trees which are
returned as JSON and rendered by the JavaScript library
d3\footnote{\url{https://d3js.org}}.

\section{Conclusions}\label{sec:conclusions}

We have outlined the design of {\sc tangos}, a system for generating
and querying databases describing halos and other objects within
cosmological simulations. We argued that the system forms a natural
`organization layer' within the simulation workflow
(Figure~\ref{fig:context-flowchart}). Sharpening this division of
responsibilities has allowed us to carry out cleaner, more focussed
and more reproducible science analyses.

{\sc Tangos} aims, above all else, to present the simplest possible
interface and so let users focus on physics. Computing efficiency is a
secondary consideration. This differs from the
simulation layer where the priority is typically to extract maximum
performance from the hardware. There is a simple reason for this
difference: the fraction of CPU resources spent on the reduction and
organization layers is, in our experience, one to two orders of
magnitude smaller than the cost of the simulation layer. On the other
hand, the fraction of human time devoted to the analysis is by far the
largest. Consequently {\sc tangos} regards human attention as the most
constrained resource.

An effective organization layer has a much broader range of
responsibilities (Figure \ref{fig:context-flowchart}) than a static
database. The modularity of {\sc tangos} allows these diverse needs ---
ranging from analysis parallelization to data organization and query
optimization --- to be satisfied by near-independent sub-packages
(Figure \ref{fig:internal-organisation}).  Despite this separation,
the sub-packages work together coherently, enabling a range of
benefits:

\begin{itemize}
\setlength\itemsep{0.9em}
\item Queries are expressed in language that reflects scientific
  intent -- but are also fast, as they are translated into carefully
  optimized joins executed by industry-standard database libraries
  (Sections \ref{sec:queries-without-sql} and
  \ref{sec:relation-finding}). The resulting interface is sufficiently
  simple that we have been able to use it with undergraduate classes,
  allowing them to undertake projects quantifying the relative role of
  mergers and smooth accretion in building the halo population.

\item We have been able to abstract away from 
  halos to a broader class of objects in the database, including black
  holes and tracked Lagrangian regions.  Especially in combination
  with the links system (Section~\ref{sec:relation-finding}), this
  allowed us to perform simultaneous analysis of the evolution of
  galaxies and their black holes \cite[e.g.][]{Romulus17,diCintio17}
  as well as dynamical analyses of stellar subpopulations tracked over
  time \citep{Pontzen15Milking}.

\item When building databases, users can implement analysis that is
  small, readable and devoid of any I/O. We have found
  this aspect particularly useful when combining analyses of multiple
  collaborators into final science results
  \citep[e.g.][]{Pontzen17GM}, a task that would previously involve
  wrangling multiple files in different formats.

\item Parallelization of the database-building process is simple to
  apply (no analysis code needs to be changed; Section
  \ref{sec:parall-strat}). This enabled us to scale from a package
  initially focussed on zoom simulations to one capable of ingesting
  state-of-the-art uniform volume runs \citep{Romulus17}.

\item Queries can be executed from Python or from a web browser
  (Section \ref{sec:webserver}). This opens up an agile mode of
  working where we perform rapid exploration of our data within a
  browser, before generating final versions of science plots using
  Python.
\end{itemize}

\vspace*{0.1cm}
{\sc Tangos} is an ongoing project and we hope that it will benefit
from broader involvement. \changed{While the major development focus
  has been on enabling us to efficiently answer questions related to
  galaxy formation}, the modular architecture should enable
improvement and growth in multiple directions, coordinated by Github's
management tools\footnote{\url{http://github.com}} and quality-assured
by Travis\footnote{\url{http://travis-ci.org}} automated testing. We
are keen to discover whether the simulation community finds {\sc
  tangos} helpful and we value all forms of feedback.

\section*{Acknowledgments}
\changed{We thank the anonymous referee for a very helpful report.}
We are grateful for discussions with and beta testing by Lauren
Anderson, Tobias Buck, Iryna Butsky, Akaxia Cruz, Arianna Di Cintio,
Ben Keller, Matthew Orkney, Martin Rey, Angelo Ricarte and Ray
Sharma. Arianna Di Cintio also suggested the name {\sc tangos},
although AP and MT take the blame for reverse-engineering it to an
acronym. We are grateful to Peter Pontzen for assistance with the
tutorial video. AP was funded by the Royal Society. MT was partially
supported by NSF award AST-1514868 and gratefully acknowledges support
from the YCAA Prize Postdoctoral Fellowship. This work used the DiRAC
Complexity system, operated by the University of Leicester IT
Services, which forms part of the STFC DiRAC HPC Facility
(\url{www.dirac.ac.uk}). This equipment is funded by BIS National
E-Infrastructure capital grant ST/K000373/1 and STFC DiRAC Operations
grant ST/K0003259/1. DiRAC is part of the National
E-Infrastructure. This work was partially enabled by funding from the
UCL Cosmoparticle Initiative.

\bibliography{paper.bib}

\begin{thebibliography}{}
\expandafter\ifx\csname natexlab\endcsname\relax\def\natexlab#1{#1}\fi

\bibitem[{{Bernyk} {et~al.}(2016){Bernyk}, {Croton}, {Tonini}, {Hodkinson},
  {Hassan}, {Garel}, {Duffy}, {Mutch}, {Poole}, \& {Hegarty}}]{BernykTAO}
{Bernyk}, M., {Croton}, D.~J., {Tonini}, C., {et~al.} 2016, \apjs, 223, 9

\bibitem[{{Di Cintio} {et~al.}(2017){Di Cintio}, {Tremmel}, {Governato},
  {Pontzen}, {Zavala}, {Bastidas Fry}, {Brooks}, \&
  {Vogelsberger}}]{diCintio17}
{Di Cintio}, A., {Tremmel}, M., {Governato}, F., {et~al.} 2017, \mnras, 469,
  2845

\bibitem[{{Hearin} {et~al.}(2017){Hearin}, {Campbell}, {Tollerud}, {Behroozi},
  {Diemer}, {Goldbaum}, {Jennings}, {Leauthaud}, {Mao}, {More}, {Parejko},
  {Sinha}, {Sip{\"o}cz}, \& {Zentner}}]{HaloTools17}
{Hearin}, A.~P., {Campbell}, D., {Tollerud}, E., {et~al.} 2017, \aj, 154, 190

\bibitem[{{Kauffmann} \& {White}(1993)}]{KauffmannWhite93}
{Kauffmann}, G., \& {White}, S.~D.~M. 1993, \mnras, 261,
  doi:10.1093/mnras/261.4.921

\bibitem[{{Lemson} \& {Virgo Consortium}(2006)}]{LemsonMillenniumSimDb}
{Lemson}, G., \& {Virgo Consortium}, t. 2006, ArXiv Astrophysics e-prints,
  astro-ph/0608019

\bibitem[{{McAlpine} {et~al.}(2016){McAlpine}, {Helly}, {Schaller}, {Trayford},
  {Qu}, {Furlong}, {Bower}, {Crain}, {Schaye}, {Theuns}, {Dalla Vecchia},
  {Frenk}, {McCarthy}, {Jenkins}, {Rosas-Guevara}, {White}, {Baes}, {Camps}, \&
  {Lemson}}]{McalpineEagleDb}
{McAlpine}, S., {Helly}, J.~C., {Schaller}, M., {et~al.} 2016, Astronomy and
  Computing, 15, 72

\bibitem[{{Nelson} {et~al.}(2015){Nelson}, {Pillepich}, {Genel},
  {Vogelsberger}, {Springel}, {Torrey}, {Rodriguez-Gomez}, {Sijacki}, {Snyder},
  {Griffen}, {Marinacci}, {Blecha}, {Sales}, {Xu}, \&
  {Hernquist}}]{Nelson15Illustris}
{Nelson}, D., {Pillepich}, A., {Genel}, S., {et~al.} 2015, Astronomy and
  Computing, 13, 12

\bibitem[{{Pontzen} {et~al.}(2015){Pontzen}, {Read}, {Teyssier}, {Governato},
  {Gualandris}, {Roth}, \& {Devriendt}}]{Pontzen15Milking}
{Pontzen}, A., {Read}, J.~I., {Teyssier}, R., {et~al.} 2015, \mnras, 451, 1366

\bibitem[{{Pontzen} {et~al.}(2013){Pontzen}, {Ro{\v s}kar}, {Stinson}, \&
  {Woods}}]{pynbody13}
{Pontzen}, A., {Ro{\v s}kar}, R., {Stinson}, G., \& {Woods}, R. 2013, {pynbody:
  N-Body/SPH analysis for python}, Astrophysics Source Code Library, , ,
  ascl:1305.002

\bibitem[{{Pontzen} \& {Tremmel}(2018)}]{tangos104}
{Pontzen}, A., \& {Tremmel}, M. 2018, Zenodo, doi:10.5281/zenodo.1248829

\bibitem[{{Pontzen} {et~al.}(2017){Pontzen}, {Tremmel}, {Roth}, {Peiris},
  {Saintonge}, {Volonteri}, {Quinn}, \& {Governato}}]{Pontzen17GM}
{Pontzen}, A., {Tremmel}, M., {Roth}, N., {et~al.} 2017, \mnras, 465, 547

\bibitem[{{Pontzen} {et~al.}(2008){Pontzen}, {Governato}, {Pettini}, {Booth},
  {Stinson}, {Wadsley}, {Brooks}, {Quinn}, \& {Haehnelt}}]{Pontzen08DLA}
{Pontzen}, A., {Governato}, F., {Pettini}, M., {et~al.} 2008, \mnras, 390, 1349

\bibitem[{{Riebe} {et~al.}(2013){Riebe}, {Partl}, {Enke}, {Forero-Romero},
  {Gottl{\"o}ber}, {Klypin}, {Lemson}, {Prada}, {Primack}, {Steinmetz}, \&
  {Turchaninov}}]{Multidark}
{Riebe}, K., {Partl}, A.~M., {Enke}, H., {et~al.} 2013, Astronomische
  Nachrichten, 334, 691

\bibitem[{{Tremmel} {et~al.}(2017){Tremmel}, {Karcher}, {Governato},
  {Volonteri}, {Quinn}, {Pontzen}, {Anderson}, \& {Bellovary}}]{Romulus17}
{Tremmel}, M., {Karcher}, M., {Governato}, F., {et~al.} 2017, \mnras, 470, 1121

\bibitem[{{Turk} {et~al.}(2011){Turk}, {Smith}, {Oishi}, {Skory}, {Skillman},
  {Abel}, \& {Norman}}]{yt11}
{Turk}, M.~J., {Smith}, B.~D., {Oishi}, J.~S., {et~al.} 2011, \apjs, 192, 9

\end{thebibliography}

\end{document}